\documentclass[aps,prd,superscriptaddress,showpacs,nofootinbib,notitlepage,onecolumn]{revtex4-1}
\usepackage{graphicx,subfigure,amsmath,amsfonts,amssymb,multirow,mathrsfs}
\usepackage[colorlinks,linkcolor=blue,citecolor=blue,urlcolor=blue]{hyperref}

\def\apj{ApJ~}                 % Astrophysical Journal
\def\apjl{ApJL~}                % Astrophysical Journal, Letters
\def\jcap{J Cosmology Astropart Phys~}
\def\mnras{MNRAS~}             % Monthly Notices of the RAS
\def\prc{Phys~Rev~C~}        % Physical Review C
\def\prd{Phys~Rev~D~}        % Physical Review D
\def\prl{Phys~Rev~Lett~}    % Physical Review Letters
\def\nat{Nature~}              % Nature
\def\nphysa{Nucl~Phys~A~}   % Nuclear Physics A
\def\physrep{Phys~Rep~}   % Physics Reports
\renewcommand\apjl{ApJL}

\begin{document}

\title{Phase transition and nuclear symmetry energy from neutron star observations: Constraints in light of PSR J0614--3329}

\author{Shao-Peng Tang}
\affiliation{Key Laboratory of Dark Matter and Space Astronomy, Purple Mountain Observatory, Chinese Academy of Sciences, Nanjing 210033, People's Republic of China}
\author{Yong-Jia Huang}
\affiliation{Key Laboratory of Dark Matter and Space Astronomy, Purple Mountain Observatory, Chinese Academy of Sciences, Nanjing 210033, People's Republic of China}
\affiliation{RIKEN Interdisciplinary Theoretical and Mathematical Sciences Program (iTHEMS), RIKEN, Wako 351-0198, Japan}
\author{Yi-Zhong Fan}
\email[Corresponding author.~]{yzfan@pmo.ac.cn}
\affiliation{Key Laboratory of Dark Matter and Space Astronomy, Purple Mountain Observatory, Chinese Academy of Sciences, Nanjing 210033, People's Republic of China}
\affiliation{School of Astronomy and Space Science, University of Science and Technology of China, Hefei, Anhui 230026, People's Republic of China}
\date{\today}

\begin{abstract}
The possible occurrence of a first-order hadron-quark phase transition in neutron-star interiors remains an open question. Whether such a transition can be directly tested with improved observations is a key challenge. Here, we incorporate the latest constraints, especially a new NICER radius measurement for PSR J0614--3329, into a nonparametric Gaussian process equation of state framework that explicitly includes a first-order transition. We find a Bayes factor of $B\approx2.3$ when comparing models with and without an explicit phase transition, marginally favoring its presence. At $68\%$ credibility, the transition onset density $n_{\rm PT}$ is either below $2\,n_s$ (corresponding to masses $\lesssim1\,M_\odot$, with density jump $\Delta n\sim0.5\,n_s$) or, more prominently, above $4\,n_s$ (near the central density of the heaviest neutron star, with $\Delta n\sim3\,n_s$), where $n_s$ represents the nuclear saturation density. In addition, by using symmetry-energy expansion at low densities ($<1.1\,n_s$), we infer a slope parameter $L=40.2^{+19.3}_{-14.3}$ MeV, in good agreement with nuclear-experiment values. Intriguingly, $L$ correlates positively with the radius difference between $1.4\,M_\odot$ and $2.0\,M_\odot$ stars.
\end{abstract}
\maketitle

\section{Introduction} \label{sec:intro}
Neutron stars (NSs) serve as natural laboratories for probing the equation of state (EOS) of cold, dense matter under extreme conditions. Much progress in constraining the EOS has been driven by multimessenger NS observations in recent years, e.g., Refs.~\citep{2020NatAs...4..625C, 2020Sci...370.1450D, 2022Natur.606..276H, 2022PhRvX..12a1058A, 2023NatCo..14.8352P, 2023PhRvL.130k2701N, 2023ApJ...955..100G, 2023SciBu..68..913H, 2024NatAs...8..328T}. The detection of gravitational waves from the binary NS merger GW170817 provided novel constraints on the tidal deformability of NSs \citep{2018PhRvL.121p1101A, 2019PhRvX...9a1001A}, while x-ray timing observations by the Neutron star Interior Composition Explorer (NICER) have yielded simultaneous mass-radius measurements for several pulsars (e.g., PSR J0030+0451 and PSR J0740+6620) \citep{2024ApJ...961...62V, 2024ApJ...974..294S}. These observations, combined with the existence of massive $\sim2\,M_\odot$ pulsars \citep{2010Natur.467.1081D, 2021ApJ...915L..12F}, have placed constraints on the stiffness of the NS EOS \citep{2020ApJ...892...55J, 2020PhRvD.101l3007L, 2020ApJ...893L..21R}.

Nuclear experiments and theory constrain the EOS at lower densities, especially the symmetry energy and its density dependence. Many analyses based on, e.g., nuclear masses, giant dipole resonances, and heavy‐ion collisions have converged on the symmetry energy slope $L \sim 50$ MeV \citep{2017ApJ...848..105T, 2020PhRvL.125t2702D}. In contrast, the Lead Radius Experiment-II (PREX-II) reported a neutron-skin thickness in ${}^{208}$Pb that implies a rather large $L$, an inference that appears inconsistent with other empirical determinations \citep{2021PhRvL.126q2502A, 2021PhRvL.126q2503R}. Meanwhile, perturbative quantum chromodynamic (pQCD) calculations at asymptotically high density provide rigorous boundary conditions that, under the requirements of thermodynamic stability and causality, can be extrapolated down toward NS‐relevant densities, generally favoring a relatively soft core in massive stars \citep{2020NatPh..16..907A, 2022PhRvL.128t2701K, 2023ApJ...950..107G, 2023SciBu..68..913H, 2024ApJ...974..244T, 2025arXiv250513691F}.

An outstanding question is whether a first-order phase transition (FOPT), e.g., to deconfined quark matter, takes place in NS interiors, and if so, at what density and with what strength. Many studies have devoted considerable effort to this problem, e.g., Refs.~\citep{2019PhRvL.122f1102B, 2019PhRvD..99h3014H, 2019PhRvL.122f1101M, 2019PhRvD..99j3009M, 2020PhRvD.102l3023B, 2020PhRvL.124q1103W, 2020ApJ...904..103M, 2021PhRvC.103c5802X, 2022PhRvL.129r1101H, 2022PhRvR...4b2054Y, 2023PhRvD.108i4014B, 2023PhRvD.108d3013E, 2023PhRvD.107d3005L, 2023PhRvD.107h3023Z, 2024PhRvC.110d5802G, 2024PhRvD.109j3008P, 2024PhRvD.110d3017Z, 2025ApJ...983...17H, 2025JCAP...02..002L, 2025arXiv250217859Y}.
A strong phase transition could lead to observable effects such as a sharp radius reduction in the NS mass-radius ($M$-$R$) relation \citep{1998PhRvD..58b4008L, 2018EPJA...54...28C, 2019arXiv190602522B}. Therefore, it is expected the FOPT might be directly verified from a list of well-measured NS observations \citep{2022PhRvD.105l3015P, 2023MNRAS.524.3464M, 2025arXiv250500194L, 2025PhRvD.111g4026L}. Previous studies using parametrized EOS models and NS observations have argued against the existence of a strong FOPT at relatively low densities \citep{2020ApJ...894L...8C, 2021PhRvD.103f3026T}. In a model-agnostic way, \citet{2023SciBu..68..913H} further revealed that a strong FOPT (i.e., $c_{\rm s}^2\rightarrow 0$) can only take place at the sufficiently-high densities of $\sim 5n_{s}$ (see their Fig.~2). More recently, nonparametric approaches incorporating explicit FOPT have been applied to updated data, including chiral effective field theory ($\chi$EFT) constraints at low density \citep{2024PhRvD.110g1502K}. These analyses suggest that a phase transition occurring at $2$-$3\,n_s$ is disfavored by the multimessenger NS data. 

Motivated by the above developments and by the recent NICER measurement of PSR J0614--3329 \citep{2025arXiv250614883M}, which favors a relatively small radius, we present a new analysis using an updated EOS framework. Specifically, we generate an ensemble of EOSs via a Gaussian process (GP) conditioned on low-density (for $n \le 1.1\,n_s$) EOSs constructed from the symmetry-energy expansion and including an FOPT at high densities. We then perform a joint Bayesian inference using four NICER pulsar observations (PSR J0437--4715, PSR J0030+0451, PSR J0740+6620, and PSR J0614--3329), the gravitational-wave constraints from GW170817, a maximum-mass prior, and pQCD constraints at high densities. With this comprehensive dataset, we update the EOS constraints, the NS $M$-$R$ relation, and related bulk properties. We also examine in detail the inferred phase transition parameters and symmetry energy parameters, highlighting several notable correlations among them.

\section{Methods} \label{sec:methods}
Our EOS model builds upon our previous works: one that combined a nuclear empirical expansion with a piecewise-parametric EOS \citep{2021PhRvD.104f3032T}, and another that implemented a Gaussian process approach for NS matter \citep{2024PhRvD.109h3037T}. Here we summarize the necessary details. First, in many theoretical models of cold, beta-equilibrated nuclear matter, the energy per nucleon at a given baryon density $n$ can be approximated by a quadratic expansion in isospin asymmetry, as follows:
\begin{equation}
\frac{E_{\rm nuc}}{A}(n,x) \simeq \frac{E_{\rm SNM}}{A}(n) + S_2(n)\,(1-2x)^2\,,
\end{equation}
where $x$ is the proton fraction. The symmetric nuclear matter part is expanded around saturation density $n_s$ as 
\begin{equation}
\frac{E_{\rm SNM}}{A}(n) = E_0(n_s) + \frac{K_0}{2}\left(\frac{n-n_s}{3n_s}\right)^2\,,
\end{equation}
and the symmetry energy is expanded as 
\begin{equation}
S_2(n) = S_v + L\left(\frac{n-n_s}{3n_s}\right) + K_{\rm sym}\left(\frac{n-n_s}{3n_s}\right)^2\,.
\end{equation}
Using these expansions up to $1.1\,n_s$, we generate a low-density EOS whose total energy density and pressure are
\begin{equation}
\begin{aligned}
\varepsilon(n,x) &= \varepsilon_e+n\left[ \frac{E_{\rm nuc}}{A}+xm_{\rm p}c^2+(1-x)m_{\rm n}c^2 \right], \\
p(n,x) &= n \frac{\partial \varepsilon(n,x)}{\partial n}-\varepsilon(n,x), \\
\end{aligned}
\end{equation}
where $m_{\rm p,n}$ are the proton and neutron rest masses, $\varepsilon_e$ is the energy density of a degenerate relativistic electron gas, and the proton fraction $x$ is determined by minimizing the total energy at a given $n$. 
We then match this EOS to the crust model at lower densities \citep{1971NuPhA.175..225B, 1973NuPhA.207..298N}. The crust-core transition density is identified by the vanishing of the effective incompressibility of npe($\mu$) matter at beta equilibrium and charge neutrality conditions \citep{2007PhRvC..76b5801K, 2007PhR...442..109L, 2018ApJ...859...90Z}. In our Bayesian framework, we sample the incompressibility coefficient $K_0$ from a normal distribution $\mathcal{N}(240,\,30^2)$ truncated to the interval $[210,\,270]$ MeV, the symmetry energy $S_v$ at saturation from $\mathcal{N}(31.7,\,3.2^2)$ truncated to $[22,\,44]$ MeV, the slope parameter $L$ from a uniform distribution on $[20,\,120]$ MeV, and the symmetry‐energy curvature $K_{\rm sym}$ from a uniform distribution on $[-400,\,100]$ MeV, while fixing the binding energy at saturation to $E_0(n_s)=-15.9$ MeV.

Beyond $1.1\,n_s$, we extend the EOS using a conditioned Gaussian process \citep{2019PhRvD..99h4049L, 2020PhRvD.101f3007E, 2020PhRvD.101l3007L, 2023ApJ...950..107G, 2025arXiv250507677L}. Specifically, we describe the sound speed via an auxiliary variable 
\begin{equation}
\phi(n) \equiv -\ln\!\Big(\frac{1}{c_s^2(n)} - 1\Big)\,,
\end{equation}
which is treated as a multivariate Gaussian distribution. That is, $\phi(n) \sim \mathcal{N}\big(-\ln(1/\bar{c}_s^2 - 1),\, K(n,n')\big)$, with a Gaussian kernel $K(n,n') = \eta \exp[-(n-n')^2/(2l^2)]$. The three hyperparameters of this GP, namely the variance $\eta$, correlation length $l$, and mean sound-speed-squared $\bar{c}_s^2$, are drawn from hyperprior distributions: $\eta \sim \mathcal{N}(1.25,\,0.2^2)$, $l \sim \mathcal{N}(0.5\,n_s,\,(0.25\,n_s)^2)$, and $\bar{c}_s^2 \sim \mathcal{N}(0.5,\,0.25^2)$. We condition the GP on the low-density EOS segment as follows: let $\phi_{\rm nuc}$ denote the values of $\phi(n)$ computed from the nuclear expansion EOS at densities $n_{\rm nuc}\le 1.1\,n_s$. We treat these as ``data" points for the GP with an assumed variance $\sigma^2_{\phi_{\rm nuc}} = 10^{-4}$ at each $n_{\rm nuc}$. The conditioned GP then yields 
\begin{equation}
\phi_{\rm GP}^{*} \mid n_{\rm nuc}, \phi_{\rm nuc}, \sigma^2_{\phi_{\rm nuc}}, n_{\rm GP} \sim \mathcal{N}\left(\bar{\phi}_{\rm GP}^{*}, {\rm cov}(\phi_{\rm GP}^{*})\right),
\end{equation}
where
\begin{equation}
\begin{aligned}
\bar{\phi}_{\rm GP}^* &= \bar{\phi}_{\rm GP} + K(n_{\rm GP}, n_{\rm nuc})\big[K(n_{\rm nuc}, n_{\rm nuc}) \\
&+ \sigma^2_{\phi_{\rm nuc}} I\big]^{-1}(\phi_{\rm nuc} - \bar{\phi}_{\rm GP})\,, \\
{\rm cov}(\phi_{\rm GP}^{*}) &= K(n_{\rm GP}, n_{\rm GP}) - K(n_{\rm GP}, n_{\rm nuc})\big[ \\
&K(n_{\rm nuc}, n_{\rm nuc}) + \sigma^2_{\phi_{\rm nuc}} I\big]^{-1}K(n_{\rm nuc}, n_{\rm GP})\,.
\end{aligned}
\end{equation}
In the above, $\bar{\phi}_{\rm GP} = -\ln(1/\bar{c}_s^2 - 1)$, $I$ is the identity matrix, and $n_{\rm GP}$ are logarithmically spaced between $1.1\,n_s$ and $40\,n_s$ on a high-resolution grid of 2560 points. Each EOS sample is generated by first drawing a set of GP hyperparameters from the hyperpriors, then sampling a realization of $\phi_{\rm GP}^*(n)$ from the conditioned GP, and finally solving for the corresponding sound speed $c_s(n)$ and integrating to obtain the pressure $p(n)$ and energy density $\varepsilon(n)$.

To incorporate an FOPT in this framework, we replace the GP‐derived sound‐speed profile $c_s^2(n)$ by zero over the density interval $[n_{\rm PT},\,n_{\rm PT} + \Delta n]$, thereby introducing a discontinuous jump in density. This corresponds to a Maxwell-type FOPT (sharp interface, with $c_s^2\to 0$). We do not consider the alternative Gibbs-type first-order transition, which would involve a mixed phase and a nonvanishing sound speed. Here $n_{\rm PT}$ (the transition onset) is drawn uniformly from $[1.1,\,10]\,n_s$ and $\Delta n$ (the density jump) uniformly from $[0,\,8]\,n_s$. For each EOS sample, we impose a prior that $n_{\rm PT}$ must be less than the central density $n_{\rm TOV}$ of the maximum-mass configuration (so that any phase transition lies within the stable branch of NSs). We also discard any EOS sample where $n_{\rm PT} + \Delta n > 10\,n_s$. Additionally, we constrain the maximum nonrotating NS mass $M_{\rm TOV}$ to lie between $2\,M_\odot$ and $3\,M_\odot$.

We perform a Bayesian inference incorporating several key observational and theoretical constraints. (i) NICER pulse-profile modeling has provided radius measurements for four rotation-powered millisecond pulsars: PSR J0437--4715 with $R = 11.36^{+0.95}_{-0.63}$~km ($68\%$ credible interval) \citep{2024ApJ...971L..20C}, PSR J0030+0451 with $R = 11.71^{+0.88}_{-0.83}$~km based on the ST+PDT modeling \citep{2024ApJ...961...62V} (with alternative models such as PDT-U being disfavored \citep{2024ApJ...966...98L}), PSR J0740+6620 with $R = 12.49^{+1.28}_{-0.88}$~km \citep{2024ApJ...974..294S}, and most recently PSR J0614--3329 with $R = 10.29^{+1.01}_{-0.86}$~km for a mass of $1.44^{+0.06}_{-0.07}\,M_\odot$. (ii) The gravitational-wave observations of GW170817 provide constraints on the tidal deformability of NSs \citep{2018PhRvL.121p1101A, 2019PhRvX...9a1001A}. (iii) We include a prior on the maximum mass informed by population studies (e.g., the marginalized posterior for $M_{\rm max}$ from \citealt{2024PhRvD.109d3052F}). (iv) We apply theoretical constraints from pQCD calculations at high density (around $10\,n_s$), which place bounds on the EOS stiffness \citep{2023ApJ...950..107G, 2023SciBu..68..913H}. For each EOS realization, when a strong FOPT produces a disconnected ``twin‐star" branch in the mass-radius relation, we include likelihood contributions from both stable branches in our comparison to observations.

\section{Results} \label{sec:results}
\begin{figure}
    \centering
    \includegraphics[width=0.5\textwidth]{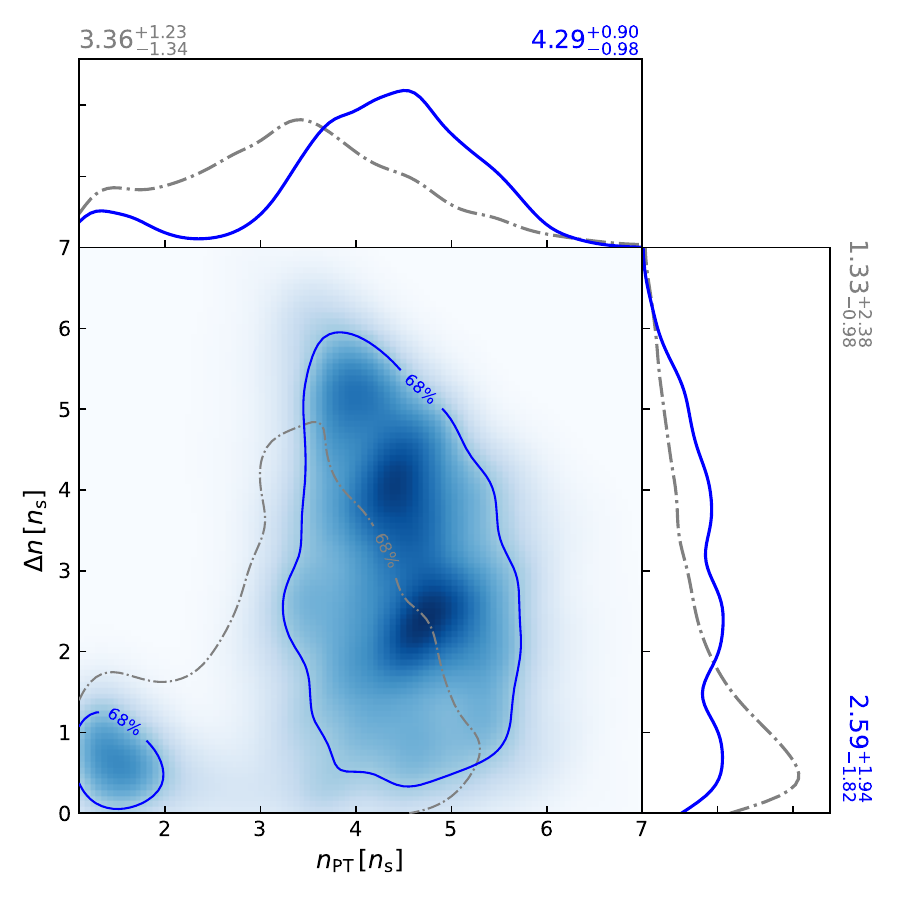}
    \caption{Prior (gray) and posterior (blue) joint distributions of the phase transition onset density $n_{\rm PT}$ and the density jump $\Delta n$.}
    \label{fig:npt-dn}
\end{figure}

\begin{figure}
    \centering
    \includegraphics[width=0.5\textwidth]{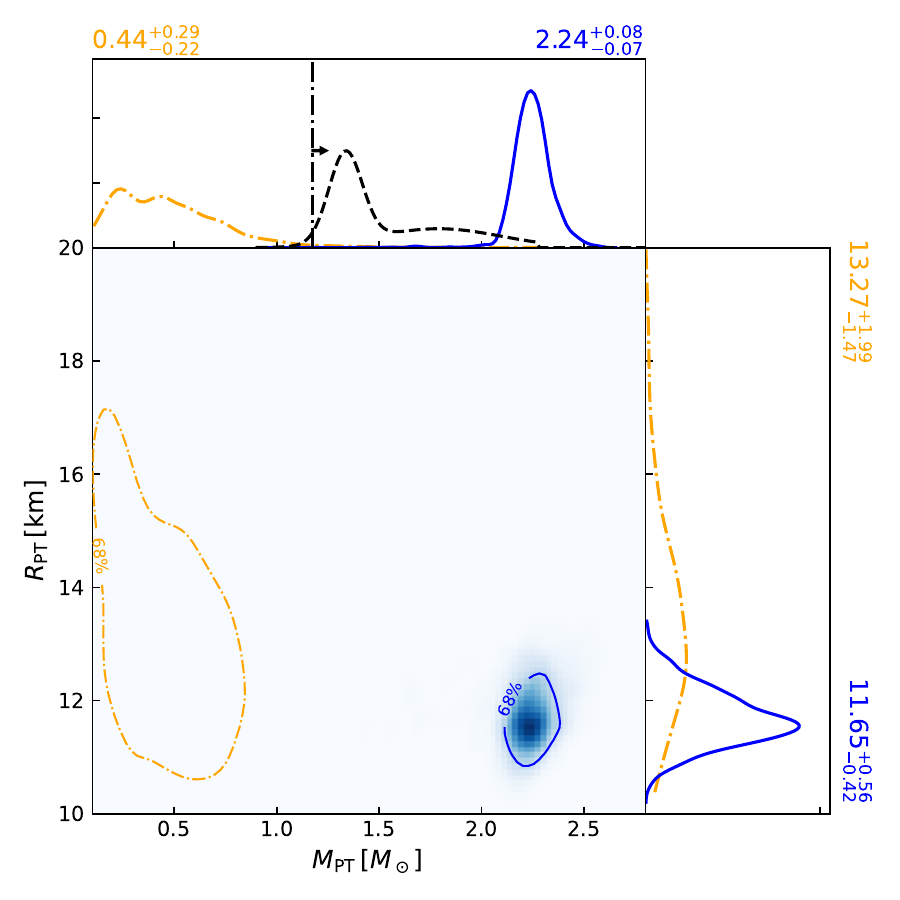}
    \caption{Mass-radius distributions for NSs whose central densities correspond to the low (orange) and high (blue) modes of the phase‐transition onset density in Fig.~\ref{fig:npt-dn}. The black dashed curve shows the observed Galactic NS mass distribution, and the vertical dot-dash line marks the lowest reliably measured mass, that of the companion to PSR J0453+1559.}
    \label{fig:mpt-rpt}
\end{figure}

\begin{figure}
    \centering
    \includegraphics[width=0.5\textwidth]{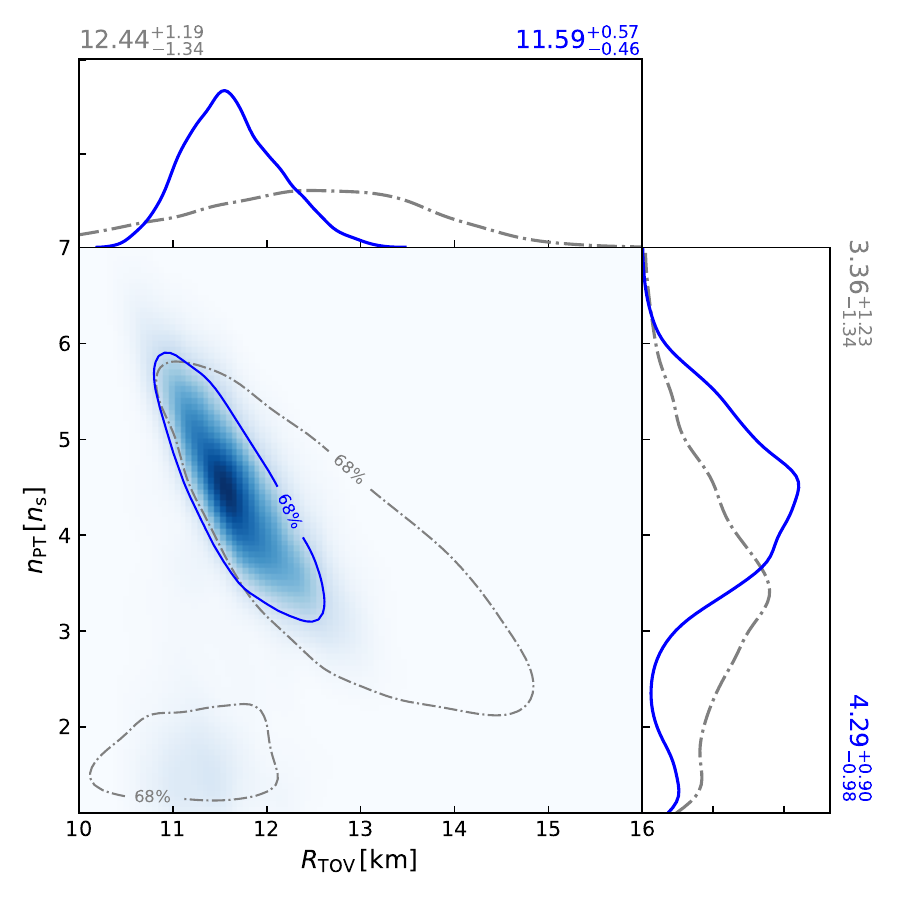}
    \caption{Prior (gray) and posterior (blue) joint distributions of $n_{\rm PT}$ and the radius of the maximum-mass NS ($R_{\rm TOV}$).}
    \label{fig:npt-rtov}
\end{figure}

\begin{figure}
    \centering
    \includegraphics[width=0.5\textwidth]{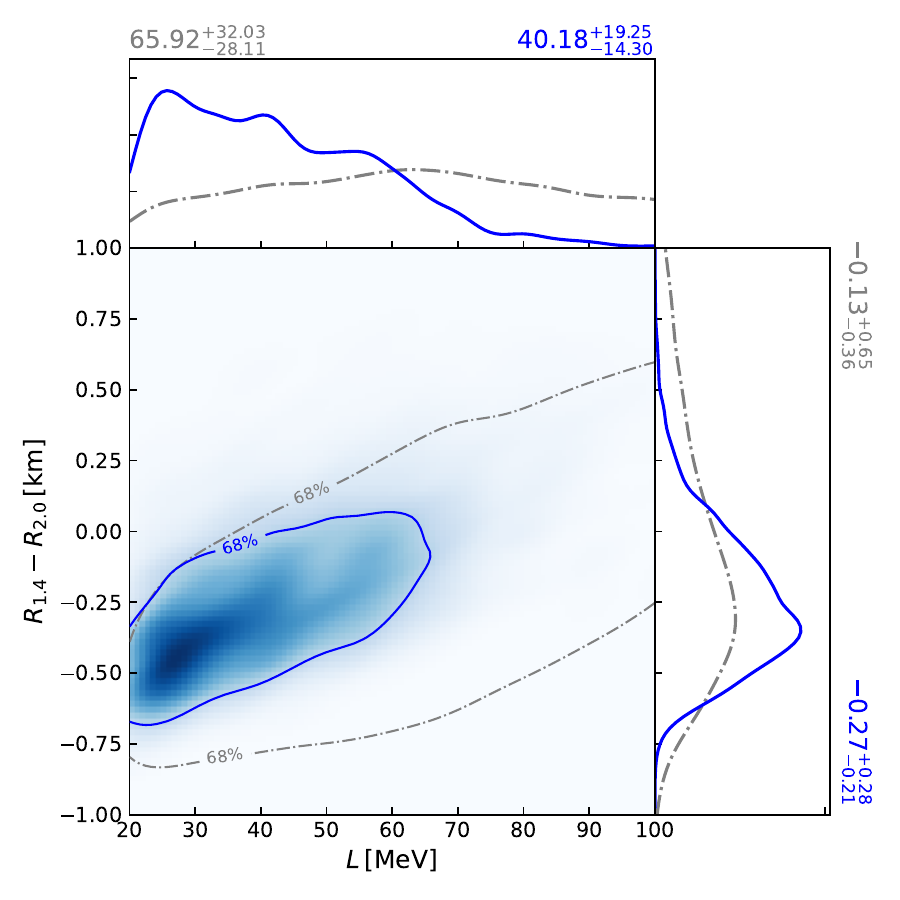}
    \caption{Prior (gray) and posterior (blue) joint distributions of the symmetry energy slope $L$ and the radius difference $R_{1.4} - R_{2.0}$.}
    \label{fig:l-dr}
\end{figure}

Figure~\ref{fig:npt-dn} compares the prior and posterior probability distributions for the phase transition parameters $n_{\rm PT}$ and $\Delta n$. We see that the updated constraints strongly disfavor a transition onset in the range $2-3\,n_s$. Instead, the posterior for $n_{\rm PT}$ is pushed toward higher densities (peaking around $4-5\,n_s$, with a small probability for $n_{\rm PT}<2\,n_s$), and the allowed density jump is limited to $\Delta n \lesssim 6\,n_s$. These results are generally consistent with recent findings by \citet{2023SciBu..68..913H} and \citet{2024PhRvD.110g1502K}, which also found that a first-order transition at $2$-$3\,n_s$ is unlikely. Our results therefore suggest that, if an FOPT occurs, its onset most likely lies in the higher‐density regime, although the uncertainty in $n_{\rm PT}$ remains substantial.

Since a strong FOPT might be directly observable (e.g, a sharp radius reduction), we examine the NS masses at which the transition starts. Figure~\ref{fig:mpt-rpt} shows the mass-radius distributions for neutron stars whose central densities coincide with the two posterior modes of the phase‐transition onset: the low‐density mode ($n_{\rm PT}<2.5\,n_s$, orange) and the high‐density mode (blue). The low‐density branch falls below $\sim1\,M_\odot$ (so lower than the lightest NS PSR J0453+1559 \citep{2015ApJ...812..143M}), while the high‐density branch lies near the maximum‐mass configuration. Both regimes lie well outside the bulk of the observed Galactic neutron‐star mass distribution (black dashed curve, \citep{2020PhRvD.102f3006S,2024PhRvD.109d3052F}), indicating that stars which sample these phase‐transition densities are exceedingly rare. Therefore, standard mass-radius measurements of typical pulsars are unlikely to directly probe the phase‐transition parameter space favored by our analysis.

Intriguingly, we find a positive correlation between the phase transition onset density and the radius of the maximum-mass NS ($R_{\rm TOV}$). As shown in Fig.~\ref{fig:npt-rtov}, EOS samples with higher $n_{\rm PT}$ tend to yield larger values of $R_{\rm TOV}$. In our earlier work, we suggested that the radius of the most massive NS could be essential for constraining high-density EOS \citep{2024PhRvD.109h3037T}. The present results reinforce this idea: if $R_{\rm TOV}$ could be determined in a good accuracy, for instance, via postmerger gravitational-wave signals \citep{2019PhRvD.100j4029B} or supramassive NS-originated black holes \citep{2024ApJ...960...67T}, it would help pinpoint the density at which a possible transition occurs. 

We also examine correlations involving the symmetry energy parameters and bulk properties of NSs. Figure~\ref{fig:l-dr} illustrates the relation between the symmetry energy slope $L$ and the radius difference between a canonical $1.4\,M_\odot$ NS and a $2.0\,M_\odot$ NS ($\Delta R = R_{1.4} - R_{2.0}$). We find that these quantities are positively correlated: EOS models with a larger $L$ generally predict a larger $\Delta R$. This implies that more rapidly decreasing radii with increasing mass (i.e., a negative $\Delta R$) are associated with lower values of $L$. Therefore, improving the precision of radius measurements for both $\sim1.4\,M_\odot$ and $\sim2.0\,M_\odot$ stars (for example, continued NICER observations of PSR J0437--4715 and PSR J0740+6620) would further constrain $L$.

\begin{figure}
    \centering
    \includegraphics[width=0.5\textwidth]{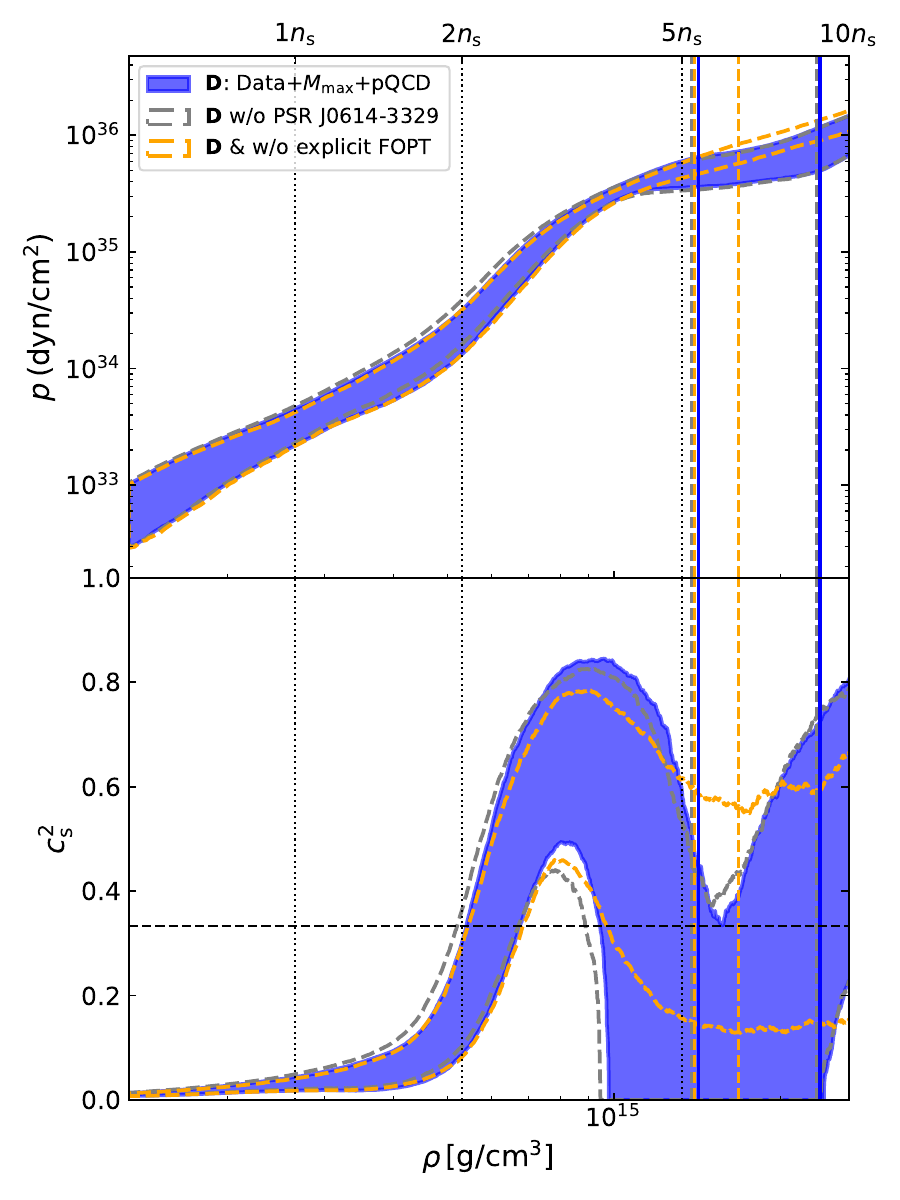}
    \caption{$68.3\%$ credible ranges for the pressure (as a function of $n$; top) and the squared sound speed $c_s^2$ (bottom), with (blue) and without (gray) the inclusion of PSR J0614--3329. The orange lines show results from EOS models without an explicit FOPT.}
    \label{fig:eos}
\end{figure}
\begin{figure}
    \centering
    \includegraphics[width=0.5\textwidth]{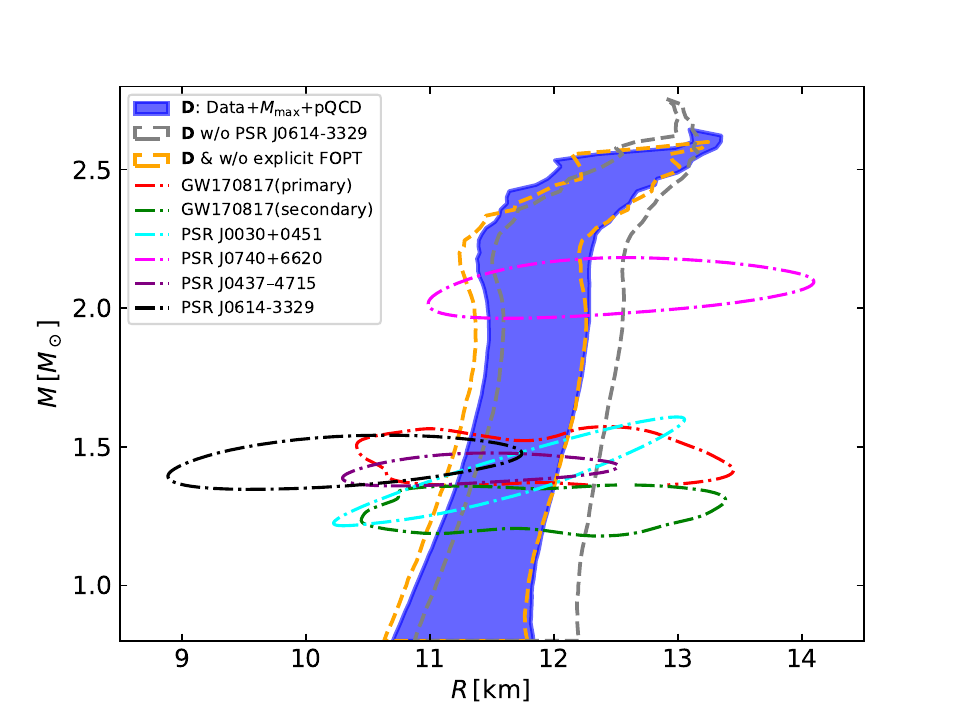}
    \caption{$68.3\%$ credible neutron star mass-radius intervals, compared to observational radius measurements. Blue curves and bands denote results including PSR J0614--3329, while gray curves correspond to results without PSR J0614--3329. The orange lines show results from EOS models without an explicit FOPT.}
    \label{fig:mr}
\end{figure}

Figures~\ref{fig:eos} and \ref{fig:mr} illustrate the impact of incorporating the new PSR J0614--3329 measurement on the EOS and $M$-$R$ posteriors. In Fig.~\ref{fig:eos}, we show the $68.3\%$ credible bands for the pressure as a function of density and for the corresponding sound speed $c_s^2(n)$. The inclusion of J0614 (blue bands) leads to a slightly softer EOS around $2\,n_s$, compared to the previous constraints (gray bands). In particular, the pressure at intermediate densities now increases more gradually (whereas the previous analysis favored a somewhat more rapid stiffening around $2$-$3\,n_s$). The behavior of $c_s^2(n)$ still exhibits a pronounced peak, and the inferred phase transition tends to occur at relatively high densities, consistent with the $n_{\rm PT}$ posterior. 

As a comparison, we also analyze an alternative EOS ensemble constructed without an explicit FOPT. Under the same dataset, the no‐FOPT posterior agrees with the FOPT case up to densities of $\sim4\,n_s$, but becomes systematically stiffer at higher densities, and its $68.3\%$ lower bound on $c_s^2(n)$ never approaches zero. We compute the Bayes factor between the FOPT and no‐FOPT hypotheses and find $B_{\rm FOPT/no\text{-}FOPT} \approx 2.3$, indicating weak (i.e., not statistically significant) preference for the inclusion of an FOPT under standard model‐comparison criteria.

Figure~\ref{fig:mr} shows the reconstructed NS mass-radius relation ($68.3\%$ credible region) alongside measured NS radius datasets. We find that adding the PSR J0614--3329 constraint shifts the overall radius estimates toward smaller values (blue region) by roughly a few hundred meters in comparison to the result without J0614 (gray region). Notably, the updated median radius for a $1.4\,M_\odot$ NS is now in closer alignment with the lower-radius mode inferred from GW170817 analyses. We also observe that the high-mass end of the $M$-$R$ curve tends to bend toward slightly larger radii (visible as a subtle upturn of the blue band for $M \gtrsim 2.25\,M_\odot$).
%This characteristic could serve as a diagnostic feature of a phase transition if future observations can map this part of the mass-radius relation.

\begin{figure*}
    \centering
    \includegraphics[width=0.75\textwidth]{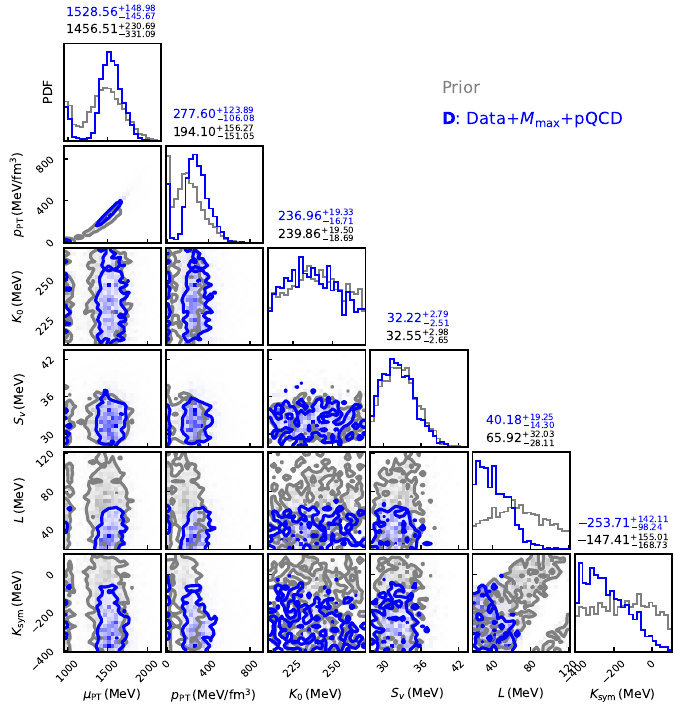}
    \caption{Prior (gray) and posterior (blue) distributions of selected EOS parameters.}
    \label{fig:params}
\end{figure*}

\begin{figure*}
    \centering
    \includegraphics[width=0.75\textwidth]{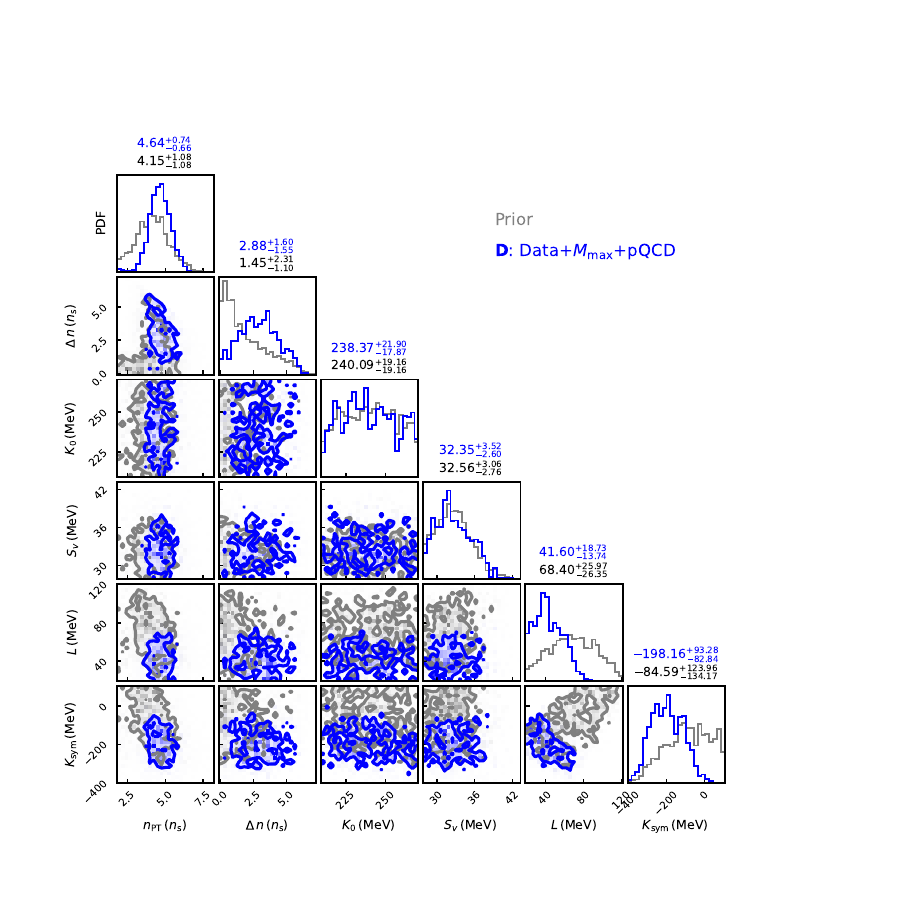}
    \caption{Prior (gray) and posterior (blue) distributions of EOS parameters when the symmetry-energy expansion is extended to $1.85\,n_s$.}
    \label{fig:newparam}
\end{figure*}

\begin{figure*}
    \centering
    \includegraphics[width=0.75\textwidth]{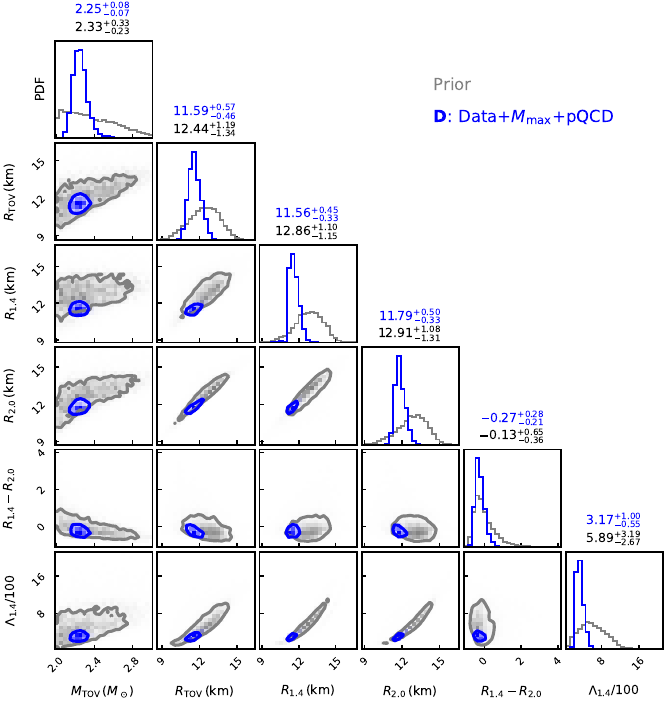}
    \caption{Prior (gray) and posterior (blue) distributions of neutron star bulk properties.}
    \label{fig:bulk}
\end{figure*}

In Fig.~\ref{fig:params}, we present the prior vs. posterior distributions for several EOS parameters. In addition to the constraints on $n_{\rm PT}$ and $\Delta n$ discussed above, we find that the symmetry energy slope $L$ is tightly constrained by the data to $40.18_{-14.30}^{+19.25}$~MeV, which is smaller than (though still consistent with) that summarized in Ref.~\citep{2021Univ....7..182L}. The posterior probability for higher $L$ values is significantly suppressed compared to the broad prior. Likewise, the symmetry energy curvature $K_{\rm sym}$ is pulled toward negative values, with only a small probability for positive $K_{\rm sym}$ remaining. The above results are obtained using the symmetry-energy expansion truncated at $1.1\,n_s$, as described in Sec.~\ref{sec:methods}. This choice follows recent work that conditions GP-based EOSs on $\chi$EFT up to $1.1\,n_s$ \citep{2024PhRvD.110g1502K}. For our model, this low cutoff keeps the expansion within its controlled domain and allows us to probe low-onset densities; however, it offers weaker constraints on the symmetry-energy parameters. We therefore repeat the analysis with the expansion extended to $1.85\,n_s$ (as adopted in our earlier parametric-EOS study \citep{2021PhRvD.104f3032T}). In that case we raise the prior lower bound on the onset density to $1.85\,n_s$ to confine the expansion to the hadronic regime and to prevent the phase transition from falling within the expansion interval. As expected (see Fig.~\ref{fig:newparam}), extending to the higher density tightens the constraints on the symmetry-energy parameters, particularly $L$ and $K_{\rm sym}$, which become $41.6_{-13.7}^{+18.7}\,\mathrm{MeV}$ and $-198_{-83}^{+93}\,\mathrm{MeV}$, respectively, while leaving the inferences for other parameters essentially unchanged relative to the $1.1\,n_s$ setup.

Figure~\ref{fig:bulk} shows the distributions for key NS bulk properties. The maximum nonrotating mass $M_{\rm TOV}$ is inferred to be $\sim2.25\,M_\odot$, and the corresponding radius is about $R_{\rm TOV}\approx11.6$~km, confirming our previous conclusions \citep{2024PhRvD.109d3052F, 2025arXiv250421408T} and indicating the robustness of this constraint. The radii of canonical-mass NSs (e.g., $R_{1.4}$ and $R_{2.0}$) are also well constrained, with uncertainties on the order of only a few hundred meters. Notably, the posterior for the radius difference $\Delta R = R_{1.4} - R_{2.0}$ is centered at a negative value (indicating $R_{2.0} > R_{1.4}$), whereas the prior allowed a wider range including large positive $\Delta R$ values that are now strongly disfavored. A negative $\Delta R$ is a potential indicator of a phase transition, as it reflects an EOS that becomes softer at higher densities (causing the most massive stars to have relatively smaller radii). Finally, the tidal deformability of a $1.4\,M_\odot$ NS, $\Lambda_{1.4}$, is constrained to be on the order of $\sim300$, in line with the tighter radius constraints.

\section{Summary} \label{sec:summary}
Thanks to an expanding array of multimessenger observations, our understanding of the neutron star EOS continues to improve. The NICER mission has now provided radius measurements for four pulsars, and although each individual measurement still has substantial uncertainty, collectively these data have significantly reduced the allowed EOS parameter space. In this work, we revisited the question of a possible phase transition inside the NSs, using flexible EOS models and incorporating the latest observational inputs (including the new NICER result for PSR J0614--3329).
We extend our previous study \citep{2023SciBu..68..913H} by considering an explicit FOPT in a model-agnostic way. Comparing EOS ensembles with and without an explicit PT yields a Bayes factor of 2.3, marginally favoring the PT hypothesis. While the result is generally consistent with the conclusion of Ref.~\citep{2023SciBu..68..913H} that $c_s^2$ approaches zero near the central density of the most massive NS, the inferred PT points suggest two scenarios: low-density PT at densities below $2\,n_s$ and more prominently, high-density PT at densities above $\sim 4\,n_s$. We further study the critical mass at which the PT occurs at the center of NSs. In the low‐density scenario, the transition onset corresponds to masses of $0.44^{+0.29}_{-0.22}\,M_\odot$, with the upper bound falling below the precisely measured mass of the lightest observed neutron star (the companion to PSR J0453+1559 at $1.174 \pm 0.04\,M_\odot$ \citep{2015ApJ...812..143M}). In the high‐density scenario, the PT nearly coincides with the maximum‐mass configuration, with a mass of $2.24^{+0.08}_{-0.07} M_{\odot}$ ($M_{\rm TOV} = 2.25^{+0.08}_{-0.07} M_{\odot}$) and a radius of $11.65^{+0.56}_{-0.42}$ km ($R_{\rm TOV} = 11.59^{+0.57}_{-0.46}$ km). Extremely large density jumps are ruled out by the pQCD constraints.
Furthermore, we find that $n_{\rm PT}$ is well correlated with $R_{\rm TOV}$. This suggests that a precise determination of $R_{\rm TOV}$ in the future (for example, via postmerger gravitational-wave signatures or other novel methods) could provide critical insight into the presence of a phase transition in NS matter.
We have also obtained tighter constraints on the symmetry energy parameters. The slope parameter $L$ is constrained to be $40.18^{+19.25}_{-14.30}$~MeV in our analysis. Interestingly, we observe that $L$ correlates with the difference in radii between $1.4\,M_\odot$ and $2.0\,M_\odot$ NSs. This correlation means that improved radius measurements for pulsars of different masses (e.g., continued observations of PSR J0437--4715 at $\sim1.4\,M_\odot$ and a future NICER or enhanced X-ray Timing and Polarimetry (eXTP) measurement of a heavy pulsar around $2\,M_\odot$) could further tighten the constraints on $L$. Such astrophysical measurements of $L$ would be especially valuable given the current tension between the PREX-II neutron-skin result and other nuclear experimental results.
Finally, we examined the effect of the new PSR J0614--3329 data on the inferred EOS. We found that including this latest NICER measurement shifts the preferred mass-radius relation toward smaller radii by an average $\sim 0.3$ km, and slightly softens the pressure at intermediate densities relative to our previous results. Many bulk NS properties (such as $R_{\rm TOV}$, $R_{1.4}$, and $R_{2.0}$) are now determined with uncertainties on the order of only a few hundred meters in the $68\%$ confidence interval. Importantly, the inclusion of the new data and the explicit phase-transition modeling does not significantly change our previously obtained $M_{\rm TOV}$, underscoring the robustness of the maximum-mass constraint near $2.2$-$2.3\,M_\odot$ \citep{2013PhRvD..88f7304F, 2024PhRvD.109d3052F, 2025arXiv250421408T}.

Given the results above, the possible observable effect for a strong FOPT (e.g., a sharp radius reduction) is favored to lie outside the mass-radius range accessible to NS observations. Current multimessenger constraints suggest that it is hard to directly confirm the FOPT from NS observations alone, even with precise mass-radius measurements. Future confirmation of FOPT is expected to be achieved by incorporating both NS observations for the cold EOS and its finite-temperature properties constrained by nuclear experiments. For this purpose, we report the chemical potential for the PT, $\mu_{\rm PT}$, at zero temperature, which also shows two peaks, located at $\sim$ 1000 and $\sim$ 1500 MeV.

Looking ahead, as multidisciplinary data accumulate from both space telescopes for NS observations (like the continuing operation of NICER and next-generation missions under development, e.g., eXTP \citep{2019SCPMA..6229502Z, 2019SCPMA..6229503W, 2025arXiv250608104L}) and terrestrial nuclear experiments, such as the Relativistic Heavy Ion Collider, the Large Hadron Collider, the Super Proton Synchrotron, the Facility for Antiproton and Ion Research, the Nuclotron-based Ion Collider Facility, and high-intensity heavy-ion accelerator facilities \citep{2020PhR...853....1B, 2022APPSB..32...35Z, 2024PrPNP.13404080S}. It is promising to resolve the remaining questions regarding possible phase transitions at both zero and finite temperature. These advances would be connected by numerical simulations for binary NS mergers \citep{2019PhRvL.122f1102B, 2019PhRvL.122f1101M, 2020PhRvL.124q1103W, 2024PhRvD.109j3008P, 2022PhRvL.129r1101H, 2024arXiv240709446H}, potentially shedding light on the location of the conjectured critical end point \citep{2025arXiv250610065E, 2025arXiv250620418C}.

\begin{acknowledgments}
This work is supported by the National Natural Science Foundation of China under Grants No. 12233011 and No. 12303056, the Project for Young Scientists in Basic Research (No. YSBR-088) of the Chinese Academy of Sciences, and the Postdoctoral Fellowship Program (No. GZC20241915) of the China Postdoctoral Science Foundation. 
\end{acknowledgments}

\section*{DATA AVAILABILITY}
The data that support the findings of this article are openly available (\url{https://doi.org/10.5281/zenodo.17149145}).

\bibliographystyle{apsrev4-1}
\bibliography{refs}

\end{document}